# A Survey of Virtualization Technologies With Performance Testing

*Joshua S. White*
josh@securemind.org

*Adam W. Pilbeam*
adam@pilbeamengineering.com

***Abstract** -- **Virtualization has rapidly become a go-to technology for increasing efficiency in the data center. With virtualization technologies providing tremendous flexibility, even disparate architectures may be deployed on a single machine without interference. Awareness of limitations and requirements of physical hosts to be used for virtualization is important. This paper reviews the present virtualization methods, virtual computing software, and provides a brief analysis of the performance issues inherent to each. In the end we present testing results of KVM-QEMU on two current Multi-Core CPU Architectures and System Configurations.***

***Keywords-Virtual Computing, Isolation, Security***

## I. Introduction

Virtualization is a mechanism permitting a single physical computer to run sets of code independently and in isolation from other sets. This is typically done in one of four ways. Each method provides the code to be run with an abstraction later to communicate with, thus limiting interaction with the physical host. In this way, virtualization has the potential to allow a single physical host the ability to act as multiple hosts that operate independently. The benefit may be seen at once: under-utilized equipment may have its utility increased in a way that will keep applications from having a negative impact on each other. With performance control mechanisms, the independent virtual machines can be kept from robbing others of physical host resources (i.e. CPU, RAM, &c.)

Virtualization has been heralded as a breakthrough solution to a number of CND (Computer Network Defense) and systems consolidation issues. ([1], [10]) However, there has not been a comprehensive evaluation of the Virtualization Technologies available and a systematic approach to evaluating them.

Previous benchmarks on various virtual machine software and hardware has shown conflicting results. [2, 5, 6, 12] This is due to a number of factors including revisions of the Virtualization Software used, the hardware configurations, CPU architectures, etc. Some previous results have shown that OpenVZ demonstrated the lowest overhead and high system performance. However OpenVZ is limited in its hardware support. [2] Yet another set of research has shown that KVM performed well in single guest implementations but has limited scalability when compared to Xen. [6] However both of these previous papers are limited due to the age of the Virtualization software used. Virtualization has come a long way in recent years, and a more modern methodology is discussed herein.

An extensive review of the methods used in past research to test performance of different VM technologies has shown the following as the de facto testing standards. ([2], [6], [8]. [9])

- Netperf [3] tests the network performance of real and virtual machines.
- IOZone [4] tests file system read/write performance
- Kernel Compile Time, a timed kernel compile using standard parameters.
- Nbench [7] CPU, FPU, and Memory Benchmarks

Thus these programs are used to further the point that proper performance evaluation must be done on a case by case basis before choosing a specific virtualization technology. Though this should not be thought of as the only factor in choosing, additional thought must go into the selection of virtual machine subsystem, licensing, and so on.

## II. Virtualization Techniques

Not all virtualization schemes are created equal. Some methods may use different, methods to implement the virtual subsystems and provide features that other methods may not be able to provide. Three primary implementations of virtualization are: ([2], [12], [15], [16])

- Emulation
- Native Virtualization
- Paravirtualization
- Operating System Level Virtualization
- Resource Virtualization
    - Storage Virtualization
- Application Virtualization

### A. Emulation (EM)

EM is a virtualization method in which a complete hardware architecture may be created in software. This software is able to replicate the functionality of a designated hardware processor and associated hardware systems. This method provides tremendous flexibility in that the guest OS may not have to be modified to run on what would otherwise be an incompatible architecture. Emulation features tremendous drawbacks in performance penalties as each instruction on the guest system must be translated to be understood by the host system. This translation process is extremely slow compared to the native speed of the host, and therefore emulation is really only suitable in cases where speed is not critical, or when no other virtualization technique will serve the purpose.

| Product | Type | License | Highlights | Performance |
|---|---|---|---|---|
| Bochs | Emulator | Open Src. | Supports Guest Debugging | Very Slow |
| QEMU | Emulator / Native | Open Src. | Large HW Support | Close to 100% |
| VMWare | Native Virtualization | Commercial | Mature, has support | Close to 100% |
| VirtualBox | Native Virtualization | Dual License | RDP Support | Close to 100% |
| Xen | Paravirtualization | Open Src. | On-the-fly Guest Migration | 100% |
| Open VZ | OS Level Containers | Open | Resource partitioning | 100% |
| User Mode Linux | Paravirtualization | Open | Stable for Linux | Close to 100% |

*Table 1: Existing Virtualization Software Overview Analysis [5]*

### B. *Native Virtualization (NV)*

NV techniques are another method for providing virtualized guests on a host system. This method does not seek to provide to the guests any hardware that is different from that of the host, thus any guest software must be host compatible. A software called a 'hypervisor' serves to translate the commands of the guest OS to the host hardware. A hypervisor may oversee several guest systems on a single host, and hypervisors are found in several virtualization methods. NV lies as a middle ground between full emulation, and paravirtualization, and requires no modification of the guest OS to enhance virtualization capabilities. This compromise allows for an increase in speed (and indeed with hardware acceleration it can be very fast), but potential performance degradation can exist in an environment where the instructions are relying more heavily on the emulated actions rather than the direct hardware access portions of the hypervisor.

### C. *Paravirtualization (PV)*

*PV* is the method wherein a modified guest OS is able to speak directly to the hypervisor. This reduces translation time and overhead as the symbiotic relationship of the two is more efficient. It does not necessarily require the guest to look for the same hardware as the host, since the guest has limited communication with the host. However, the number of operating systems available for guest usage may be limited depending on the availability of paravirtualization-aware driver kits that must be used to ensure compatibility with the hypervisor. [2]

### D. *Operating System Level Virtualization (OSLV)*

*OSLV* is the method in which an operating system kernel provides for multiple isolated user-space instances. This is not true virtualization, however it does provide the ability for user-space applications (that would be able to run normally on the host OS) to run in isolation from other software. Most implementations of this method can define resource management for the isolated instances.

### E. *Resource Vritualization (RV)*

RV is a method in which specific resources of a host system are used by the Guest OS. These may be software based resources such as domain names, certificates, etc. or hardware based such as shared storage space. This form of virtualization is largely used in the HPC (High Performance Computing) community because of it's advantages in forming a single logical computer across multiple nodes. Such a setup is beneficial when building cluster based supercomputers. In this way, the end user no longer needs to create cluster specific applications, instead the virtual machine can be treated like a single physical machine.

#### 1) *Virtual Storage (VS)*

VS is specific form of Resource Virtualization that merits it's own subcategory. VS provides a single logical disk from many different systems across a network. This disk can then be made available to Host or Guest OS's.

### F. *Application Virtualization (AV)*

AV provides smaller single application virtual machines that allow for emulation of a specific environment on a client system. For example a Java Virtual Machine allows disparate operating systems such as Windows and Linux to run the same Java program as long as they have the Java VM installed. This form of virtualization is limited in that it only provides single program isolation from the host, but is useful when testing programs out without installing them.

## III. Virtualization Hardware Support Limitation

Direct access to a hosts hardware tends to be limited for the guest machines. Instead a virtual interface driver is used to access real hardware through the virtual systems interface. In the case of KVM [11], the hypervisor is OSS (Open Source Software) which allows for additional virtual hardware to be added at the whim of any programmer with the right knowledge base. However this is one of the major drawbacks when it comes to closed solutions such as VMWare's solutions.

## IV. Virtual Computing Software

There are several software products available now for virtualization. Each one may implement its own flavor of the methods mentioned before, and each software product may have its own benefits and limitations. Table 1 describes the software products, virtualization method used, and provides performance comparison.

Current virtualization methods each consist of a trade-off to the users. Very few methods can hope to achieve same-as-host performance capabilities with an unmodified guest operating system, which may be needed in many cases, and may be unavoidable in cases where the guest OS has no available stable modifications to allow it to be virtualized (Mac OS, etc).

In these cases a performance hit must be accepted if the unmodified OS is vital to operations. At the same time, with a growing number of operating systems that can be modified to perform using paravirtualization method, options for achieving great performance are increasing while also expanding on the flflexibility of OS choice. However, some updates and upgrades to guests may lag behind the physical host (unmodified) versions if there is no dedicated development team to stay on top of the modification.

## V. Security and Virtualization

Physical computer systems work with very little interfering layers between the Operating System and the hardware. In many cases the OS talks to a Kernel which in turn commands the hardware. This leaves many openings to threats that would exploit either or both the software and hardware of the system. There are a number of projects to include items such as Mandatory Access Control to the OS and provide added security by restricting the communication of an application to the hardware. However, Virtualization provides another layer of protection by allowing un-trusted software access only to vitual hardware. Though, this does not mean that the system is impervious to threat, only that some specific threats are better mitigated through the use of Virtual Machines.

### A. *Considerations*

Much of the same security implications for servers and server operating systems continues to apply when dealing with virtualized systems. Generally, most systems implement an independent guest operating system that has login capability, remote access, etc. these must be secured the same as any physical system. Physical access is actually somewhat limited now, as the only physical interfaces to access these virtualized machines lie with the physical host. This aspect may be beneficial, however as it may be easier to secure physically one host that supports a multiplicity of services. Conversely, the potential for malicious activity is increased is physical or remote access is attained to the physical host or host OS. Thus the physical host may be considered a single point of failure from a security and reliability standpoint. This must be addressed in any policy dealing with virtualization security.

### B. *Applications*

Moving away from the security considerations of dealing with virtualization, we can furthur look at the benefits and applications of Virtualization for CND and Systems Security. There are a number of security applications where VM's (Virtual Machines) are highly desirable. ([13], [14], [15], [16])

#### 1) *Isolation*

The first of these is Isolation, which restricts the operations which the Guest VM can perform on the Hosts physical hardware. It also allows for the VM's to have restricted access to other VM's, creating a self contained virtual network.

#### 2) *Intrusion Detection*

Intrusion Detection is the second and most beneficial application of virtualization technologies. Traditional IDS (Intrusion Detection Systems) rely on either network sensors or client software. The issue that arises is that these systems either can't see the whole picture of what is going on in the client system or the software runs on the client and is therefor prone to attack. By implementing Intrusion Detection in the Host OS and interfacing it with the VMM (Virtual Machine Manager) we have an out-of-band tool that can see everything happening on the guest OS.

#### 3) *Redundancy*

The last security application that we'll discuss is that of redundancy. In a traditional server environment when a system becomes infected, it can take some time to bring that system down for a cleaning or rebuild, especially if it's critical to a mission. Virtual guests can be thought of as files in that the machines exist as images and therefore can be backed up as such. In addition trusted template systems can be build and brought up at a moments notice to fill the void of a critical server.

## VI. KVM-QEMU System Test System Configuration

For our tests we selected a relatively standard low end generic server. The following specifications denote the system that all tests were performed on.

- Dual Core AMD Athlon X2 4800+
- 4 GB of DDR-3 Memory
- 1 x 500 GB SATA-II Hard Drive for the Host Operating System

- 1 x 2 TB SATA-II Western Digital Green Hard Drive with 32 MB Cache for Guest Operating Systems
- 1 x 10/100/1000 nVidea MCP61 Integrated Ethernet Network Interface

Standard bridging of eth0 to vnet0 was done so that the guest could have direct access to the hosts physical network interface.

The host OS (Operating System) was a Base KDE Desktop install of Fedora 12. The guest was a duplication of the host install with all the same features and software selected. Virt-Manager 0.7.0-8 was used to easily manage the virtual machine. QEMU version qemu-0.10.6-9.fc12.x86_64 was used in conjunction with KVM qemu-kvm-0.10.6-9.fc12.x86_64 to provide full simulation of target architecture.

## VII. KVM-QEMU System Test Results and Analysis

The following results analysis shows the performance in a number of categories of KVM-QEMU versus the native host OS. It's worth mentioning that the Guest OS was given only 1 GB of RAM to work with while the Host OS had the full 4GB to work with. As a result the test can not be thought of as comparing to like systems, but instead a simple performance analysis.

### A. *Netperf Results*

The following plots show the results of the Netperf network throughput testing. A selection of what has been considered the most relevant tests including TCP Send File, TCP Stream and TCP MAERTS tests were selected. [17]

- TCP Stream
  - Tests the flow of data from the local netperf client to the remote netserver but does not count the handshake and connection setup time.
- TCP MAERTS
  - Tests the flow of data from the remote netserver to the local netperf client.
- TCP Send File
  - Tests the flow of data from the local netperf client to the remote netserver application. The calculation of time does not include the handshake or setup time. It is different from TCP Stream in that it uses the sendfile() function call which results in lower CPU utilization because the transfer comes directly out of the from the Filesystem Buffer.

A direct comparison can be made between the Guest and Host for network performance as the amount of available RAM that the VM has available plays no part in a direct access test such as this. Increased latency is caused by the bridging function of the host and the virtual network interface. For this test a third system on the network running an identical OS and software configuration was used. The connection between the physical systems was a 10/100 Ethernet Switch.

As evident from Plots 1 and 2 the results while somewhat staggered are actually quite close. Throughput was only slightly effected by the additional virtual layers between the OS and hardware.

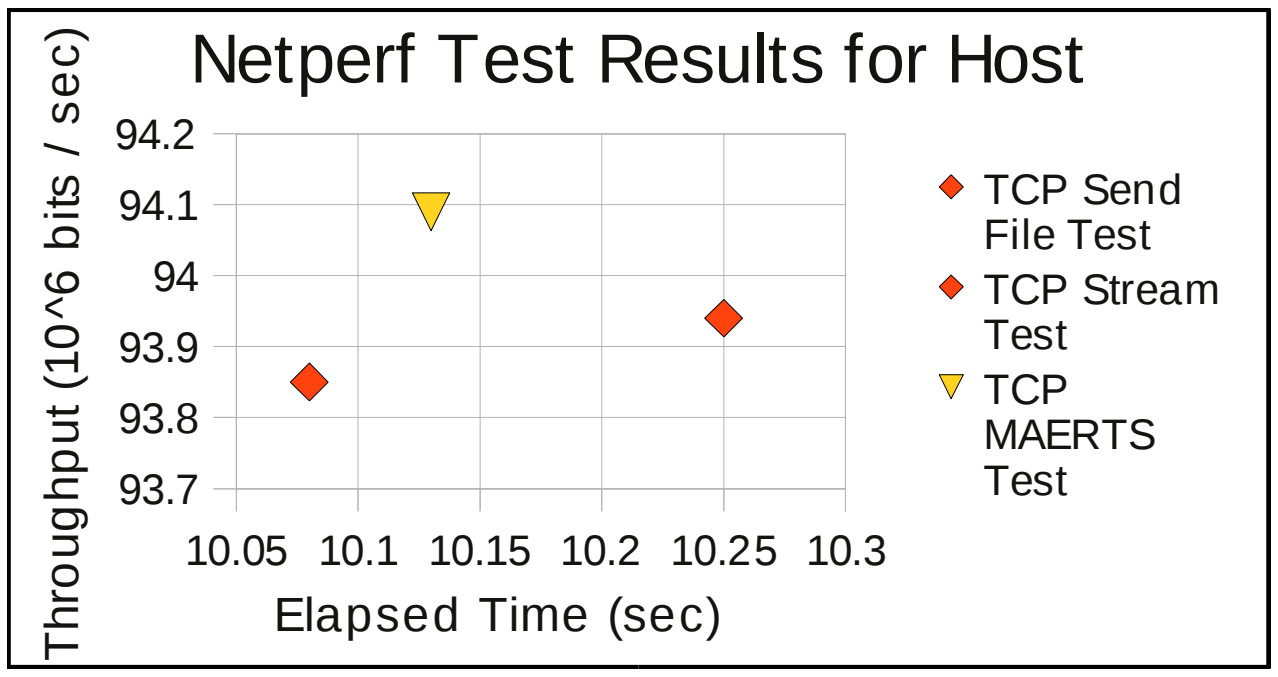

*Plot 1: Netperf Performance Results for Host System*

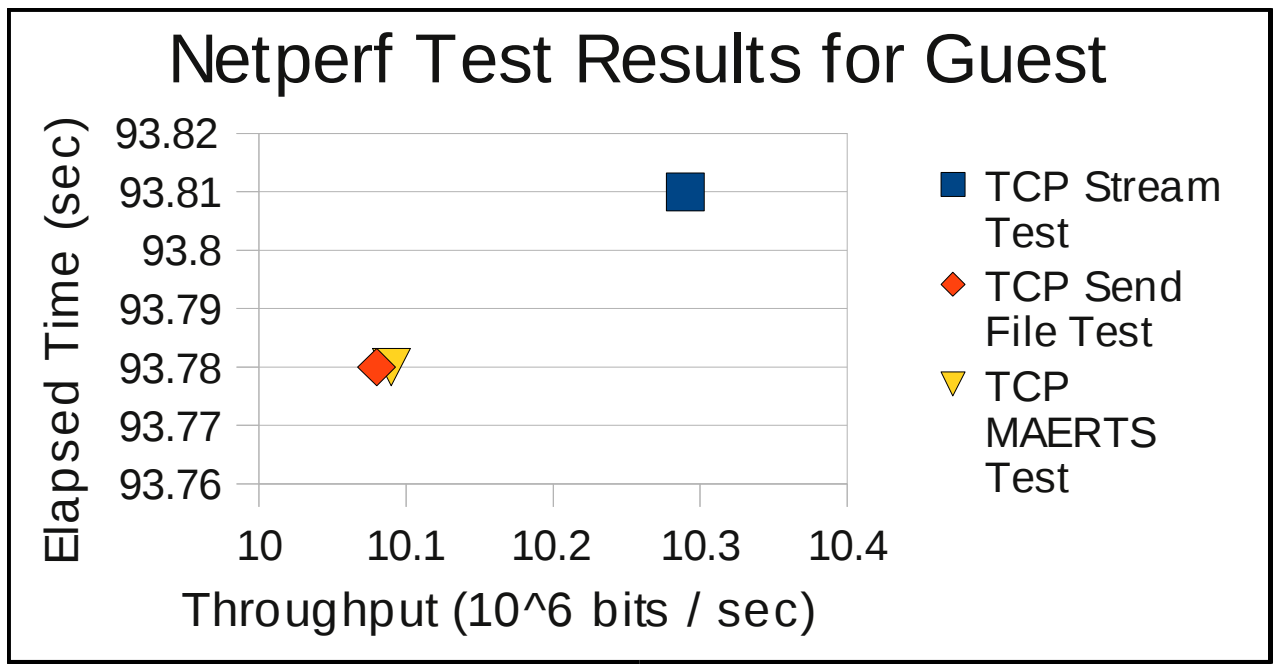

*Plot 2: Netperf Performance Results for Guest System*

### B. *IOZone Results*

Read and write to the filesystem performance measurements were made using IOZone. Plots 3 and 4 show the results for a 64 KB data size.

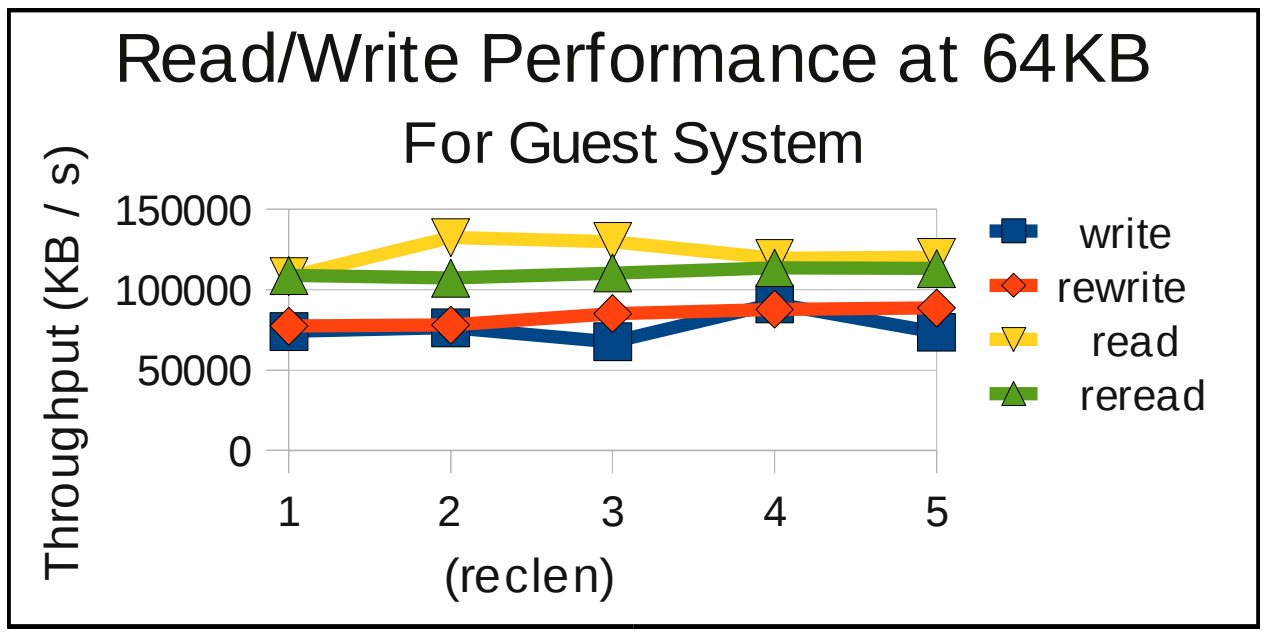

*Plot 3: IOZone Read/Write 64 KB Results for Guest System*

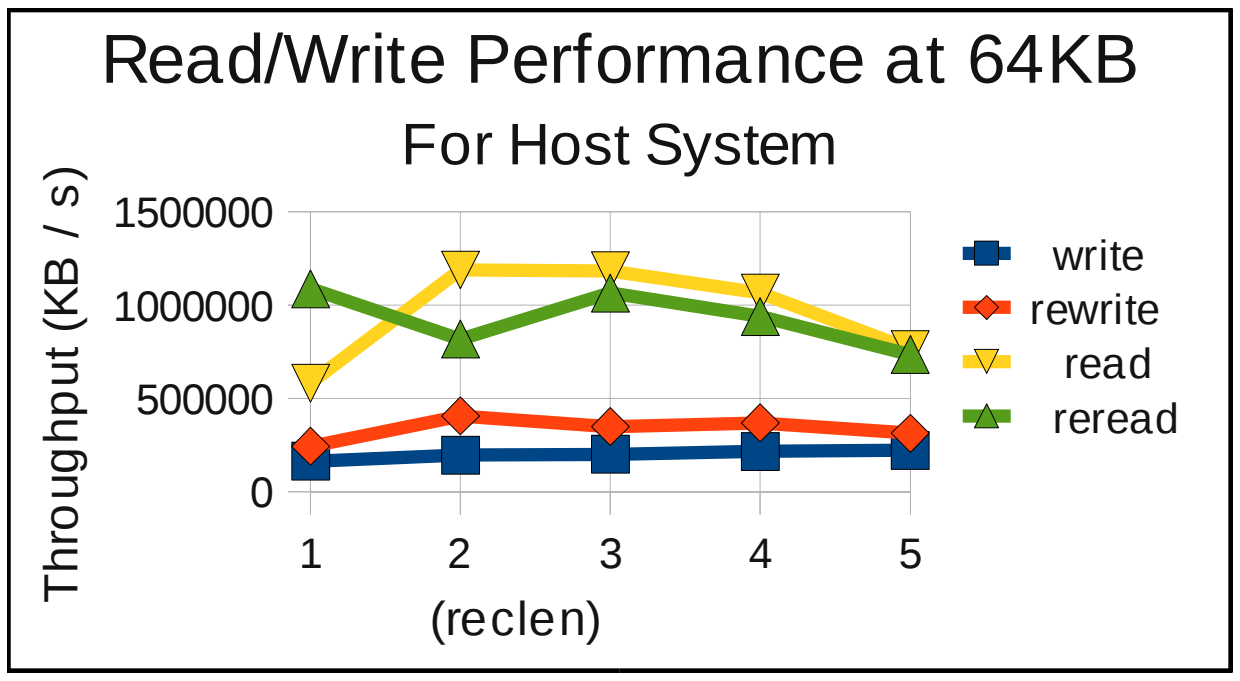

*Plot 4: IOZone Read/Write 64 KB Results for Host System*

The plots show similar results between the host and guest system but this is harder to directly compare because the virtual machine environment is writing to a file-system within a file, while the host environment is writing to a file-system spread across a physical disk. Still from the perspective of performance impact, again we can see that there's no detrimental impact to using the Guest VM.

### *C. Kernel Compile Results*

The kernel compile time measurements can not be directly compared because the Guest and Host had such a large difference in available RAM. However the results are still very promising for the VM even with its memory limitation. Table 2 shows the results of a generic compile of the 2.6.33.4 Linux kernel with default options selected.

| | **Host Time** | **Guest Time** |
|---|---|---|
| **real** | 90m27.646s | 121m27.752s |
| **usr** | 65m19.017s | 70m4.279s |
| **sys** | 14m16.185s | 43m11.199s |

*Table 2: Kernel Compile Time Measurements For Host and Guest*

## VIII. Discussion and Future Work

It has been shown that virtualization can play different roles in a computing system, be it for defense, consolidation, etc. Once the role is known a proper VM mechanism can be chosen and evaluated allowing minimal performance impact to be achieved.

Each of the virtualization methodologies presents its own benefits. In cases where the guest operating system cannot be modified for enhanced virtualization performance, only native virtualization and emulation methods may be implemented. Additionally, cross architecture virtualization is remanded to the realm of emulation only. In such cases where modification to the guest operating systems are available, the enhancements afforded by paravirtualization can and should be taken advantage of. All the mechanisms and methods must still be secured, both for each virtual machine on its own as well as the physical host. The final analysis must be then that the selection of a virtualization method for a single physical host must be based on application based policies and reassessed as new technologies become available.

It is proposed that future work be done in developing a methodology for selecting a virtualization solution based on a flow model. It is further proposed that the taxonomy presented in the paper "Virtualization: a Survey on Concepts, Taxonomy and Associated Security Issues" [15] be augmented with an available solutions list. The taxonomy combined with a selection methodology would allow for standardized approach to choosing virtual machine software solution.

## X. Authors

**Joshua S. White** (M'06) Received an AAS degree in Computer Network Technologies from Finger Lakes Community College, Canandaigua NY, and both BS and MS degrees in Telecommunications from the State University of New York Institute of Technology, Marcy NY. This author became a Member (M) of the IEEE in 2006.

He currently works as the Director of Research and Development for a small government contracting company. His experience includes WAN Protocol Analysis tools and WAN vulnerability research. He's successfully proposed two new areas of research to government agencies including WAN protection and Malware nullification.

Mr. White belongs to the ISSA (Information Systems Security Association), and is the current President of its Central New York Chapter. Additionally he is a member of the NY Infragaurd and OISF (Open Information Security Foundation).

**Adam W. Pilbeam** Received an AAS degree in Telecommunications Technology from Cayuga Community College, Auburn, NY and both BS and MS degrees in Telecommunications from the State University of New York Institute of Technology, Marcy NY.

He currently works as the lead Network Engineer for a small government contracting company. His experience includes WAN Protocol Analysis and Next Generation WAN deployment research. He is well versed in Linux Kernel modification, Linux hardening and server design/lock-down.

Mr. Pilbeam is a member in good standing of the ISSA (Information Systems Security Association), and has given numerous presentations to the Central New York Chapter.